\documentclass[aps,10pt]{revtex4}
\usepackage{amsfonts}
\usepackage{amsmath}
\usepackage{amssymb}
\usepackage{graphicx}

\input{tcilatex}

\begin{document}

\title{Mode Coupling in Quantized High Quality Films}
\author{Yiying Cheng and A. E. Meyerovich}
\affiliation{Department of Physics, University of Rhode Island, 2 Lippitt Rd., Kingston\\
RI 02881-0817}

\begin{abstract}
The effect of coupling of quantized modes on transport and localization in
ultrathin films with quantum size effect (QSE) is discussed. The emphasis is
on comparison of films with Gaussian, exponential, and power-law long-range
behavior of the correlation function of surface, thickness, or bulk
fluctuations. For small-size inhomogeneities, the mode coupling is the same
for inhomogeneities of all types and the transport coefficients behave in
the same way. The mode coupling becomes extremely sensitive to the
correlators for large-size inhomogeneities leading to the drastically
distinct behavior of the transport coefficients. In high-quality films there
is a noticeable difference between the QSE patterns for films with bulk and
surface inhomogeneities which explains why the recently predicted new type
of QSE with large oscillations of the transport coefficients can be observed
mostly in films with surface-driven relaxation. In such films with
surface-dominated scattering the higher modes contribute to the transport
only as a result of opening of the corresponding mode coupling channels and
appear one by one. Mode coupling also explains a much higher transport
contribution from the higher modes than it is commonly believed. Possible
correlations between the inhomogeneities from the opposite walls provide,
because of their oscillating response to the mode quantum numbers, a unique
insight into the mode coupling. The presence of inhomogeneities of several
sizes leads not to a mechanical mixture of QSE patterns, but to the overall
shifting and smoothing of the oscillations. The results can lead to new,
non-destructive ways of analysis of the buried interfaces and to study of
inhomogeneities on the scales which are inaccessible for scanning techniques.
\end{abstract}

\maketitle

\section{Introduction}

Progress in material technology, especially in nanofabrication, ultrathin
film deposition, ultraclean and high vacuum systems, \textit{etc., }requires
better understanding of the effect of remaining bulk inhomogeneities or
surface defects on physical processes in high-quality systems. In
high-quality systems, these remaining inhomogeneities small and smooth with
a low amplitude and a relatively large lateral scale. In some cases, such as
in ultrathin films, the lateral scale of bulk and surface inhomogeneities
can even be much larger than the film thickness. Scattering by such small,
but long-range inhomogeneities is crucial for transport in ultrathin and/or
clean systems in which the particle mean free path is comparable to the
system size.

Below we consider the effect of random, mostly large-scale, bulk, surface,
and thickness fluctuations on quantum transport in quantized quasi-$2D$
systems such as quantized flow channels, waveguides, or ultrathin metal
films. We will look at the single-particle diffusion coefficient $D$ in a
channel as a function of the particle energy and the channel width and at
the low-temperature mobility $\mu $ (conductivity $\sigma $) as a function
of the film thickness and the Fermi wavelength. The main issue is to find
how sensitive is the transport to the statistical properties of
inhomogeneities, \textit{i.e., }to the structural or thickness fluctuations
with small amplitude and large correlation radius. Here we have in mind
large-size surface steps and thickness fluctuations for ultrathin films,
slow long-range bending of fibers or films, slowly fluctuating bulk fields, 
\emph{etc. }One of the main goals is to separate the effect of the
scattering-driven mode coupling from other scattering effects.

The choice of quasi-$2D$ systems is explained by a desire to avoid
divergence of surface fluctuations and strong localization effects which are
inherent to $1D$ systems. In contrast to $1D$ systems, the randomly
fluctuating $2D$ surfaces are stable while the localization length in
systems with weak fluctuations is exponentially large.

Usual approaches to bulk and surface fluctuations are different from each
other. The bulk fluctuations are routinely described via the fluctuating
bulk potential $V\left( \mathbf{r}\right) $\ or, whenever possible, via the
scattering $T$-matrix, $T\left( \mathbf{p,p}^{\prime }\right) $. Since $%
\widehat{V}$\ and $\widehat{T}$ are tied to each other via the integral
equation,%
\begin{equation}
\widehat{T}=\widehat{V}+\widehat{T}\widehat{G}\widehat{V}  \label{eq1}
\end{equation}%
($\widehat{G}$\ is the Green's function), these two descriptions are, in
principle, equivalent (and, in the case of weak fluctuations - identical).
Below we assume that the bulk inhomogeneities are defined by their
scattering $\widehat{T}$-matrix and that this $\widehat{T}$-matrix is known.

The prevalent way to characterize the surface roughness or thickness
fluctuations is to use the correlation function of the surface
inhomogeneities 
\begin{equation}
\zeta \left( \mathbf{s}\right) \equiv \zeta \left( \left\vert \mathbf{s}%
\right\vert \right) =\left\langle \xi (\mathbf{s}_{1})\xi (\mathbf{s}_{1}+%
\mathbf{s})\right\rangle \equiv A^{-1}\int \xi (\mathbf{s}_{1})\xi (\mathbf{s%
}_{1}+\mathbf{s})d\mathbf{s}_{1},  \label{aa1}
\end{equation}%
where $\mathbf{s}$ gives the $2D$ coordinates along the surface, $\xi (%
\mathbf{s})$\ describes the deviation of the position of the surface in the
point with $2D$ coordinates $\mathbf{s}$ from its average position, $%
\left\langle \xi (\mathbf{s})\right\rangle =0$, and $A$ is the averaging
area. This equation\ assumes that the correlation properties of the surface
do not depend on the lateral direction.

Any transport theory should yield an explicit dependence of the transport
coefficients (the particle mean free path, diffusion or mobility
coefficients, lateral conductivity, \emph{etc.}) on the correlator of
surface inhomogeneities $\zeta \left( s\right) $ and/or the bulk scattering $%
T$-matrix. Both of these objects can be characterized by their average
amplitudes $\ell $, types of decay (exponential, power law, \emph{etc.}) at
large distances or momenta, and the correlation radii $R$ which
characterizes this decay. For the surface correlation functions $\zeta $,
parameters $\ell $ and $R$ are the average \textquotedblright
height\textquotedblright\ and \textquotedblright lateral
size\textquotedblright\ of surface inhomogeneities or thickness
fluctuations. For the bulk scattering $T$-matrix, $\ell $ is the average
amplitude in the dependence of the scattering amplitude on dimensionless
momentum $pR/\hbar $ and $R$ provides a scale for decay of the scattering
amplitude at large momentum transfers. The transport coefficients are
determined by the relation between the particle wavelength $\Lambda ,$ the
width of the channel $L$, and the correlation radius of inhomogeneities $R$.
If the fluctuations are weak, the fourth length parameter, $\ell $, (more
precisely, its square) enters as a coefficient. For example, the
conductivity $\sigma $ of degenerate fermions and the single-particle
diffusion coefficient $D$ can be parameterized as 
\begin{eqnarray}
\sigma &=&\frac{2e^{2}}{\hbar }\frac{R^{2}}{\ell ^{2}}f\left( \Lambda
,L,R\right) ,  \label{bb1} \\
D &=&\frac{\hbar }{m}\frac{R^{2}}{\ell ^{2}}d\left( \Lambda ,L,R\right) .
\label{d1}
\end{eqnarray}%
with virtually identical functions $f$ and $d$. The reason for this
commonality is that both of these transport coefficients are expressed via
the same combination of the zeroth and first angular harmonics of the
scattering probabilities.

We consider ultrathin systems with quantized motion across the film.
Scattering by surface and bulk inhomogeneities could cause coupling of the
otherwise distinct quantum modes. We will show that the qualitative behavior
of the transport coefficients is extremely sensitive to this
scattering-driven coupling of the modes which, in turn, is determined by the
long-range behavior of the structural or thickness fluctuations. Mode
coupling has already attracted considerable attention for localization \cite%
{and1} and transport, especially in the context of applications of the
random matrix theory \cite{matr1}. Though most of the applications involved
transport in systems with bulk disorder, systems with surface disorder were
also considered \cite{mar1,arm2,pon1}.

What has been mostly ignored is the sensitivity of the mode coupling and, in
the end, transport, to the type of correlation behavior of disorder. It has
been often assumed that the correlation function is short-range ($\delta $%
-type correlations, hard spheres, \emph{etc.}) without long tails. What is
more, in the case of impurities a natural assumption is that the correlation
radius $R$\ ("size"\} of the disorder is relatively small. Under these
assumptions the mode coupling is indeed featureless (though robust) and does
not lead to any striking effects which depend on the nature of disorder.
However, in high-quality quasi-$2D$ samples, it is possible to observe bulk
and surface disorder with various correlation properties and fluctuations of
different scales. In this case, as we will see below, the mode coupling and
transport can follow several distinct scenarios. These different types of
behavior are determined by the rate of decay of correlations and their scale
and not by the origin of fluctuations or nature and spectra of the particles
and waves.

Recently we predicted \cite{pon1} a new type of quantum size effect (QSE)
with huge, large-period oscillations of conductivity $\sigma \left( L\right) 
$ in films with dominant surface scattering. This surface-driven effect is
so large that a real puzzle is why this new type of QSE has not been
observed earlier in high quality films with bulk scattering. Below we will
answer this question by comparing films with bulk and surface scattering. We
will also analyze the contribution of different modes to transport. A usual
assumption is that in films with surface scattering the main contribution to
transport comes form the grazing particles - modes with the lowest quantum
numbers. We will see that the scattering-driven mode coupling makes this
assumption wrong even for the highest quality films. The last important
issue that will be studied below is the possibility of extracting
information on the quality of the film surfaces, including the buried ones,
from the non-destructive transport measurements. Interestingly, QSE in
transport can even reveal a possible correlation between random surface
inhomogeneities from the opposite walls.

\section{Transport in quantized systems}

Below we deal mostly with the conductivity of metal films and the
single-particle diffusion in quasi-$2D$ channels. QSE experiments in metal
films include conductivity \cite{conduct1,qse2}, spectroscopy \cite{spect1},
susceptibility \cite{susc1}, and STM \cite{stm1} measurements. The signature
feature of QSE in metals is a pronounced saw-tooth dependence of the lateral
conductivity on, for example, film thickness, $\sigma \left( L\right) $,
common for both bulk \cite{rr6} and surface \cite{r4} scattering. However,
the QSE experiments in metals have to overcome a difficulty which one does
not encounter in semiconductors. The period of the saw-tooth QSE
oscillations in the dependence $\sigma \left( L\right) $ is usually small,
nearly atomic, $\hbar /p_{F}$, making the saw-tooth behavior of transport
almost impossible to observe. For this reason, typical experimental objects
are lead or semimetal films such as bismuth. In contrast to this "standard"
saw-tooth effect, QSE, which is inherent to high quality films \cite{pon1},
results in smooth, large-period oscillations of $\sigma \left( L\right) $ at
relatively large values of $p_{F}L/\hbar $. This QSE could be observed in a
wider group of metals. Large-period QSE oscillations have already been
observed (see the second Ref. \cite{conduct1}); however, the experimental
details are still sketchy. This issue is also related to the long-standing
controversy on the influence of the structure of the nanoscale film on its
resistivity \cite{qse2}.

Another group of seemingly different physical applications involves the
single-particle diffusion in quantized flow channels. Though the typical
examples - quantized helium quasiparticles in ultrathin channels \cite%
{gu1,reyer1} and ultracold neutrons in gravitational traps \cite{neutron1} -
seem to be far apart from metal films, the descriptions of the transport
processes in such diverse systems are very similar to each other.

Recently, we developed a transparent semi-analytical formalism for transport
in systems with rough boundaries that allows simple uniform calculations in
a wide range of systems and for various types of roughness \cite{arm1,arm2}.
One attractive feature is that this formalism treats the scattering by
surface inhomogeneities using the same transport equation as for the bulk
imperfections and even allows one to study the interference between bulk and
surface scattering \cite{arm3}. This formalism unites earlier approaches by
Tesanovic \textit{et al } \cite{r3}, Fishman and Calecki \cite{qq17},
Kawabata \cite{kaw1}, Meyerovich and S. Stepaniants \cite{r2}, and Makarov 
\textit{et al}\cite{mak1} (for a brief review of different theoretical
approaches see Refs.\cite{arm2,arm4}). In this paper we apply this approach
to the study of the fluctuation-driven coupling of quantized modes. The
limits of applicability of our approach are discussed in detail in Refs.\cite%
{arm2,arm3}.

Since the $2D$ mobility of particles is described by essentially the same
equations as the exponent in the expression for the localization length in
films, our study also provides the dependence of the localization length on
the type of the correlation function of random surface and bulk
inhomogeneities.

The paper has the following structure. In the next Section we introduce the
transport equation and expressions for the transport coefficients. Section
IV briefly describes various types of the surface inhomogeneities and bulk $%
T $-matrices used in the computations. The results are given in Section V
and conclusions - in Section VI.

\section{Transport equation and transport coefficients}

QSE is caused by quantization of motion in the direction perpendicular to
the film, $p_{x}\rightarrow \pi j\hbar /L$, which leads to a split of the
energy spectrum $\epsilon \left( \mathbf{p}\right) $ into a set of
minibands, $\epsilon \left( p_{x},\mathbf{q}\right) \rightarrow \epsilon
\left( \pi j\hbar /L,\mathbf{q}\right) =\epsilon _{j}\left( \mathbf{q}%
\right) $. For simplicity, we consider mostly particles with a parabolic
spectrum, 
\begin{equation}
\epsilon _{j}\left( \mathbf{q}\right) =\frac{1}{2m}\left[ \left( \pi j\hbar
/L\right) ^{2}+q^{2}\right] .  \label{t1}
\end{equation}%
The extension to non-parabolic spectra is discussed in Section VE. We will
look at two similar transport problems, namely, at conductivity of
degenerate fermions,%
\begin{equation}
q_{j}\equiv q_{Fj}=\left[ 2m\epsilon _{F}-\left( \pi j\hbar /L\right) ^{2}%
\right] ^{1/2},  \label{t2}
\end{equation}%
($\epsilon _{F}=\pi ^{2}\hbar ^{2}/2m\Lambda _{F}^{2}$ is the overall Fermi
energy, $\Lambda _{F}$ is the Fermi wavelength, and $q_{j}\left( \epsilon
_{F},L\right) $ is the Fermi momentum for the miniband $j$) and at
single-particle diffusion for particles with energy $E=\pi ^{2}\hbar
^{2}/2m\Lambda ^{2}$,%
\begin{equation}
q_{j}=\left[ 2mE-\left( \pi j\hbar /L\right) ^{2}\right] ^{1/2}  \label{t3}
\end{equation}%
where $q_{j}\left( E,L\right) $ is the momentum of such a particle in the
miniband $j$. Both problems are computationally similar and reduce to almost
identical sets of linear equations \cite{arm2}.

In the case of conductivity of degenerate fermions, the transport equation
for the distribution functions $n_{j}\left( \mathbf{q}\right) ,$ 
\begin{equation}
\frac{dn_{j}}{dt}=2\pi A\sum_{j^{\prime }}\int W_{jj^{\prime }}\left[
n_{j^{\prime }}-n_{j}\right] \delta \left( \epsilon _{j\mathbf{q}}-\epsilon
_{j^{\prime }\mathbf{q}^{\prime }}\right) \frac{d^{2}q^{\prime }}{\left(
2\pi \hbar \right) ^{2}},  \label{aa3}
\end{equation}%
reduces, after standard transformations, to a set of linear equations 
\begin{eqnarray}
q_{j}/m &=&-\sum_{j^{\prime }}\nu _{j^{\prime }}\left( q_{j^{\prime
}}\right) /\tau _{jj^{\prime }},  \label{ee6} \\
\frac{2}{\tau _{jj^{\prime }}} &=&m\sum_{j^{\prime \prime }}\left[ \delta
_{jj^{\prime }}\,W_{jj^{\prime \prime }}^{\left( 0\right) }-\delta
_{j^{\prime }j^{\prime \prime }}\,W_{jj^{\prime }}^{\left( 1\right) }\right]
\notag
\end{eqnarray}%
where $n_{j}^{\left( 1\right) }=\nu _{j}\delta \left( \epsilon -\epsilon
_{F}\right) eE$ is the first angular harmonic of the distribution function $%
n_{j}\left( \mathbf{q}\right) $ at $q=q_{j}$, and $W_{jj^{\prime }}^{\left(
0,1\right) }\left( q_{j},q_{j^{\prime }}\right) $ are the zeroth and first
harmonics of $W\left( \mathbf{q}_{j}\mathbf{-q}_{j^{\prime }}\right) $\ over
the angle $\widehat{\mathbf{q}_{j}\mathbf{q}}_{j^{\prime }}$. The solution
of Eqs. $\left( \text{\ref{ee6}}\right) $ provides the conductivity of the
film: 
\begin{equation}
\sigma =-\frac{e^{2}}{3\hbar ^{2}}\sum_{j}\nu _{j}\left( q_{j}\right) q_{j}.
\label{ee7}
\end{equation}%
The matrix $\widehat{\tau }$ is diagonal when the scattering-driven mode
coupling is negligible with respect to the in-band scattering.

The single-particle diffusion is similar to the conductivity problem for
degenerate fermions. In equilibrium, a particle with energy $E$ can be in
any of $S$ accessible minibands $\epsilon _{j}\left( \mathbf{q}\right)
=\left( 1/2m\right) \left[ \left( \pi j\hbar /L\right) ^{2}+q^{2}\right] $
for which $\epsilon _{j}\left( q=0\right) \leq E$. The equilibrium
distribution function $n^{\left( 0\right) }$ is 
\begin{equation}
n^{\left( 0\right) }\left( \mathbf{q}\right) =\sum n_{j}^{\left( 0\right)
}\left( \mathbf{q}\right) =\frac{\pi }{mS}\sum \delta \left( E-\epsilon
_{j}\left( \mathbf{q}\right) \right) ,  \label{s1}
\end{equation}%
and the transport equation reduces to a set of $S$ coupled linear equations
for distributions $\nu _{j}\left( q_{j}\right) $ with momenta $q_{j}$ $%
\left( \text{\ref{t3}}\right) $ which is almost identical to Eqs. $\left( 
\text{\ref{ee6}}\right) ,\left( \text{\ref{ee7}}\right) $:%
\begin{eqnarray}
\frac{1}{Sm}q_{j}\nabla \rho &=&-\sum_{j^{\prime }}\nu _{j^{\prime }}\left(
q_{j^{\prime }}\right) /\tau _{jj^{\prime }},  \label{s2} \\
D\nabla \rho &=&-\frac{1}{m}\sum_{j=1}^{S}q_{j}\nu _{j}\left( q_{j}\right) ,
\notag
\end{eqnarray}%
where $\nabla \rho $ is the density gradient that causes the diffusion, $D$
is the single-particle diffusion coefficient, and $\widehat{\tau }$ is still
defined by Eq.$\left( \ref{ee6}\right) $. The single-particle mobility
coefficient $b$ is related to $D$ by the Einstein equation $D=bE$.

The results for the single-particle diffusion also provide the mean free
path $\mathcal{L}$ and the exponent in the expression for the localization
length $\mathcal{R}$ that describes localization caused by particle
scattering by random wall and/or bulk inhomogeneities \cite{arm2}: 
\begin{equation}
\mathcal{R}=\mathcal{L}\exp \left[ \pi mSD/\hbar \right] .  \label{loc1}
\end{equation}

\section{Surface correlation function \emph{vs.} bulk scattering amplitude}

In the case of surface or thickness fluctuations, we consider an infinite $%
2D $ channel (or film) of the average thickness $L$ with random rough
boundaries 
\begin{equation}
x=L/2-\xi _{1}(y,z),\ x=-L/2+\xi _{2}(y,z).  \label{a0}
\end{equation}%
(the walls are assumed hard with infinite potential). The inhomogeneities
are small, $\xi _{1,2}\left( y,z\right) \ll L$, and random with zero
average, $\left\langle \xi _{1}\right\rangle =\left\langle \xi
_{2}\right\rangle =0$. Their correlation function $\zeta _{ik}\left( \mathbf{%
s}\right) $ and its Fourier image $\zeta _{ik}\left( \mathbf{q}\right) $,
which is often called the power spectral density function or, in short, the
power spectrum, are defined as 
\begin{eqnarray}
\zeta _{ik}\left( \left\vert \mathbf{s}\right\vert \right) &=&\left\langle
\xi _{i}(\mathbf{s}_{1})\xi _{k}(\mathbf{s}_{1}+\mathbf{s})\right\rangle
\equiv A^{-1}\int \xi _{i}(\mathbf{s}_{1})\xi _{k}(\mathbf{s}_{1}+\mathbf{s}%
)d\mathbf{s}_{1},  \label{a1} \\
\zeta _{ik}\left( \left\vert \mathbf{q}\right\vert \right) &=&\int d^{2}s\
e^{i\mathbf{q\cdot s}}\zeta _{ik}\left( \left\vert \mathbf{s}\right\vert
\right) =2\pi \int_{0}^{\infty }\zeta _{ik}\left( s\right) J_{0}\left(
qs\right) sds  \notag
\end{eqnarray}%
where $\mathbf{s=}\left( y,z\right) $ and $\mathbf{q=}\left(
q_{y},q_{z}\right) $ are the $2D$ vectors. In homogeneous systems, the
correlation function depends only on the distance between points $\left\vert 
\mathbf{s}_{1}-\mathbf{s}_{2}\right\vert $ and not on coordinates
themselves. The correlation functions $\zeta _{11}$ and $\zeta _{22}$
describe intrawall correlations of inhomogeneities, and $\zeta _{12}=\zeta
_{21}$ - the interwall correlations. Usually, but not always, the
inhomogeneities on different walls are not correlated with each other, $%
\zeta _{12}=0$. Thus, everywhere, except for Section VF, it is assumed that $%
\zeta _{12}=0$. To avoid parameter clutter, we also assume that the
correlation parameters are the same on both walls, $\zeta _{11}=\zeta
_{22}=\zeta $. Then the effective correlator contains $2\zeta \left(
s\right) $ with $\zeta \left( s\right) $ given by equations below.

Surface inhomogeneities exhibit a wide variety of types of the correlation
functions \cite{q2,fer1}. To have a meaningful comparison, we consider the
correlation functions that involve only two characteristic parameters,
namely, the amplitude (average height) $\ell $ and the correlation radius
(average size) $R$ of surface inhomogeneities.

The most commonly used in theoretical applications is the Gaussian
correlation function, 
\begin{equation}
\zeta \left( s\right) =\ell ^{2}\exp \left( -s^{2}/2R^{2}\right) ,\ \zeta
\left( q\right) =2\pi \ell ^{2}R^{2}\exp \left( -q^{2}R^{2}/2\right)
\label{a2}
\end{equation}
including its limit for small correlation radius $R\rightarrow 0$, \textit{%
i.e.,} the $\delta $-type correlations, 
\begin{equation}
\zeta \left( s\right) =\ell ^{2}R^{2}\delta \left( s\right) /s,\ \zeta
\left( q\right) =2\pi \ell ^{2}R^{2}.  \label{a3}
\end{equation}

Sometimes, a better fit to experimental data on surface scattering is
provided by the use of the exponential correlation function 
\begin{equation}
\zeta \left( s\right) =\ell ^{2}\exp \left( -s/R\right) ,\ \zeta \left(
q\right) =\frac{2\pi \ell ^{2}R^{2}}{\left( 1+q^{2}R^{2}\right) ^{3/2}},
\label{e2}
\end{equation}%
or by the even more long-range, power-law correlators 
\begin{equation}
\zeta \left( s\right) =\frac{2\mu \ell ^{2}}{\left( 1+s^{2}/R^{2}\right)
^{1+\mu }},\ \zeta \left( q\right) =2\pi \ell ^{2}R^{2\,}\frac{\left(
qR\,\right) ^{\mu }}{2^{\mu -1}\Gamma \left( \mu \right) }K_{\mu }\left(
qR\right)  \label{ee2}
\end{equation}%
with different values of the parameter $\mu $. The asymptotic behavior of
the power spectra, \emph{i.e., }of the functions $K_{\mu }\left( qR\right) $%
, is essentially exponential. The most commonly used are the Staras function
with $\mu =1$ and the correlator with $\mu =1/2$ which has the plain
exponential power spectrum $\zeta \left( q\right) $, 
\begin{equation}
\zeta \left( q\right) =2\pi \ell ^{2}R^{2}\exp \left( -qR\right) \text{.}
\label{ee22}
\end{equation}

The last class of correlation functions covers the power-law correlators in
momentum space, 
\begin{equation}
\zeta \left( q\right) =\frac{2\pi \ell ^{2}R^{2}}{\left( 1+q^{2}R^{2}\right)
^{1+\lambda }},\ \zeta \left( s\right) =\ell ^{2}\frac{\left( s/R\,\right)
^{\lambda }}{2^{\lambda }\Gamma \left( 1+\lambda \right) }K_{\lambda }\left(
s/R\right) ,  \label{ee3}
\end{equation}%
which are exponential functions in the configuration space. The correlators
from this group include the Lorentzian in momentum space $\lambda =0$ that
was observed in Ref.\cite{fer1} and the exponential correlator $\left( \text{%
\ref{e2}}\right) $ at $\lambda =1/2$.

The constants in all these correlators are chosen in such a way that the
value of $\zeta \left( q=0\right) =2\pi \ell ^{2}R^{2}$ is the same. This
provides a reasonable basis of comparison for transport coefficients in
films with all these different types of random surfaces. Indeed, the
scattering cross-section for $q\rightarrow 0$ does not depend on the details
of short- and mid-range structure of surface inhomogeneities. Therefore, at
Fermi momenta $q_{F}\rightarrow 0$ (more precisely, at $q_{F}R\ll 1$), the
transport coefficients should be the same for all random surfaces. (The only
exception is the Lorentzian $\left( \text{\ref{eee2}}\right) $ for which $%
\zeta \left( q\right) $ diverges at small $q$; however, the Lorentzian is
non-physical anyway). Some useful analytical expressions for the angular
harmonics of these correlation functions can be found in Ref. \cite{pon1}.

To have a uniform comparison of the results throughout the paper, we will
plot numerical results for the Gaussian correlator, the power-law correlator
in configuration space $\left( \text{\ref{ee2}}\right) $ with $\mu =1$ (the
Staras function), and the power-law correlator in momentum space $\left( 
\text{\ref{ee3}}\right) $ with $\lambda =0.5$ (exponential correlator in
real space).

In the case of bulk modulation, it makes sense to start directly from the
bulk scattering amplitude $T\left( \mathbf{p,p}^{\prime }\right) $ which, in
the case of quantized films, is transformed into the matrix%
\begin{equation}
T\left( \mathbf{p,p}^{\prime }\right) =T\left( p_{x},\mathbf{q;}%
p_{x}^{\prime },\mathbf{q}\right) \rightarrow T_{jj^{\prime }}\left( \mathbf{%
q,q}^{\prime }\right) =T\left( \mathbf{q}_{j}-\mathbf{q}_{j^{\prime
}}^{\prime }\right) .  \label{t5}
\end{equation}%
For this matrix, we also looked at the Gaussian form similar to $\left( 
\text{\ref{a2}}\right) $, power law form similar to $\left( \text{\ref{ee3}}%
\right) $ with $\lambda =0.5$ (standard Debye screening with an exponent in
real space), and exponential form $\left( \text{\ref{ee2}}\right) $ with $%
\mu =1$. Another interesting possibility here would be an oscillating
function which corresponds to an alternative type of screening in metals.
However, such a function would require us to introduce two lateral length
parameters which would make a meaningful comparison between the correlation
functions impossible.

In what follows we compare the transport properties of the films $\left( 
\text{\ref{a2}}\right) $ - $\left( \text{\ref{ee3}}\right) $\ in a wide
range of film thickness $L$, correlation radius $R$, and particle wavelength 
$\Lambda =\pi /q$ (or the $2D$ particle density $N$).

The transport equation for both bulk and surface imperfections is the same,
Eqs. $\left( \text{\ref{aa3}}\right) $ - $\left( \text{\ref{s2}}\right) $ \
The only distinguishing feature is the dependence of the scattering
probabilities $W_{jj^{\prime }}\left( \mathbf{q},\mathbf{q}^{\prime }\right) 
$ on the correlation function of surface fluctuations $\zeta \left(
\left\vert \mathbf{q-q}^{\prime }\right\vert \right) $ \cite{arm2,pon1},%
\begin{equation}
W_{jj^{\prime }}\left( \mathbf{q},\mathbf{q}^{\prime }\right) =\frac{\hbar }{%
m^{2}L^{2}}\left[ \zeta _{11}+\zeta _{22}+2\zeta _{12}\left( -1\right)
^{j+j^{\prime }}\right] \left( \frac{\pi j}{L}\right) ^{2}\left( \frac{\pi
j^{\prime }}{L}\right) ^{2},  \label{rrr3}
\end{equation}%
and the $T$-matrix for the bulk imperfections, 
\begin{equation}
W_{jj^{\prime }}\left( \mathbf{q},\mathbf{q}^{\prime }\right) =\frac{2\pi }{%
\hbar }\left\vert T\left( \mathbf{q}_{j}-\mathbf{q}_{j^{\prime }}^{\prime
}\right) \right\vert ^{2}.  \label{t6}
\end{equation}

Note that in contrast to Eq. $\left( \text{\ref{rrr3}}\right) $, the
dependence of $W$ $\left( \text{\ref{t6}}\right) $ on band indices $%
j,j^{\prime }$ is generally not known explicitly and is determined by the $T$%
-matrix $\left( \text{\ref{t5}}\right) $. This dependence is the same as in
Eq. $\left( \text{\ref{rrr3}}\right) $ when the fluctuating bulk field is
essentially two-dimensional and can be factorized as%
\begin{equation}
U\left( x\right) +\left( \partial U/\partial x\right) \xi \left( \mathbf{s}%
\right)  \label{t7}
\end{equation}%
where $x$ an $\mathbf{s}$ are the coordinate across and along the film and $%
U\left( x\right) $ is the field without fluctuations. This situation, though
realistic, is by no means general. If, for example, the bulk field
fluctuates only along the film and does not change across the film, then $%
T\left( \mathbf{q}_{j}-\mathbf{q}_{j^{\prime }}^{\prime }\right) =\delta
_{jj^{\prime }}T\left( \mathbf{q}-\mathbf{q}^{\prime }\right) $. The
off-diagonal elements of $T\left( \mathbf{q}_{j}-\mathbf{q}_{j^{\prime
}}\right) $ and, therefore, the mode coupling are associated only with the
variation of the bulk field across the film. All this makes modeling of the
function $T\left( \mathbf{q}_{j}-\mathbf{q}_{j^{\prime }}\right) $ more
ambiguous than for the surface correlator. There are two ways of dealing
with this. The first one is to assume that the bulk fluctuations in
ultrathin films have the form $\left( \text{\ref{t7}}\right) $ and,
essentially, to use the expression similar to Eq. $\left( \text{\ref{rrr3}}%
\right) $ for the scattering probabilities (of course, without the
oscillating interwall term $\zeta _{12}\left( -1\right) ^{j+j^{\prime }}$).
Another approach, which is more appropriate for thicker films is to assume
that the bulk fluctuations are truly three-dimensional and are not affected
by the film boundaries except from the straightforward quantization,%
\begin{equation}
T_{jj^{\prime }}\left( \mathbf{q,q}^{\prime }\right) \equiv T\left( \pi
j\hbar /L,\mathbf{q;\pi }j^{\prime }\hbar /L,\mathbf{q}^{\prime }\right) .
\label{t8}
\end{equation}%
Then, for example, the Gaussian fluctuations in momentum space lead to the
following form of $W$:%
\begin{equation}
W_{jj^{\prime }}\left( \mathbf{q},\mathbf{q}^{\prime }\right) =\frac{8\pi
^{5}\hbar ^{5}\ell ^{2}R^{2}}{m^{2}L^{6}}\exp \left( -q^{2}R^{2}/2\right)
\exp \left[ -\frac{\pi ^{2}\left( j-j^{\prime }\right) ^{2}R^{2}}{2L^{2}}%
\right] ,  \label{t10}
\end{equation}%
where we introduced parameters $\ell $ and $R$ in such a way so that to make
the description as close as possible to the one with the Gaussian thickness
fluctuations $\left( \text{\ref{a2}}\right) $ (or with the one with the
Gaussian bulk fluctuations of the type $\left( \text{\ref{t7}}\right) $),%
\begin{equation}
W_{jj^{\prime }}\left( \mathbf{q},\mathbf{q}^{\prime }\right) =\frac{8\pi
^{5}\hbar ^{5}\ell ^{2}R^{2}}{m^{2}L^{6}}j^{2}j^{\prime 2}\exp \left(
-q^{2}R^{2}/2\right) .  \label{t11}
\end{equation}

\section{Results and discussion}

\subsection{General comments}

As it is mentioned in Introduction, the $2D$ conductivity $\sigma $\ of the
film has the dimensionality of conductance, $e^{2}/\hbar $.The conductivity
depends on the relation between three length scales - particle (Fermi)
wavelength $\Lambda _{F}=1/q_{F},$ the width of the channel $L$, and the
"lateral size" (correlation radius) of inhomogeneities $R$. The fourth
length parameter, $\ell $, is perturbative and enters conductivity as a
perturbative coefficient $1/\ell ^{2}$,%
\begin{equation}
\sigma =\frac{2e^{2}}{\hbar }\frac{R^{2}}{\ell ^{2}}f\left( L/\Lambda
_{F},R/L\right)  \label{g1}
\end{equation}%
Note, that the conductivity diverges in the limit of vanishing
inhomogeneities $\ell \rightarrow 0$\ or $R\rightarrow \infty $.

The single-particle diffusion coefficient $D$ can be parameterized in a
similar way,%
\begin{equation}
D=\frac{\hbar }{m}\frac{R^{2}}{\ell ^{2}}d\left( L/\Lambda ,R/L\right)
\label{g2}
\end{equation}%
where the dimensionless function $d$ is given by the same combination of the
zeroth and first harmonics of the scattering probability $W$,%
\begin{equation}
d\left( L/\Lambda ,R/L\right) =\frac{3}{2S}f\left( L/\Lambda ,R/L\right)
\label{g3}
\end{equation}%
and $S=\mathrm{Int}(L/\Lambda )$ is the number of quantized energy minibands
accessible to the particle with the wavelength $\Lambda $. The presence of
the step-wise function $S(L/\Lambda )$ in the denominator of Eq.$\left( \ref%
{g3}\right) $ can lead to a visible difference in shapes between functions $%
d\left( L/\Lambda ,R/L\right) $ and $f\left( L/\Lambda ,R/L\right) $: at
large $R/L$ the function $f$ is practically smooth while the function $d$
has residual singularities in the point where the number of accessible
minibands changes. These singularities can serve as useful markers that help
to separate the "standard" QSE and the QSE of Ref. \cite{pon1}. Therefore,
we will plot the results for $d\left( L/\Lambda ,R/L\right) $ more often
than for $f\left( L/\Lambda \right) $. Apart from these small-amplitude
singularities, the functions $d$ and $f$ are similar.

Below we will plot the dimensionless functions $d\left( L/\Lambda \right) $\
and $f\left( L/\Lambda _{F}\right) $ at various constant values of $R/L$.
The orders of magnitude of these functions change widely depending on the
type of surface and bulk correlators and the value of $R/L$. In order to
provide a better visual comparison between these functions, we will
normalize $d\left( x\right) $\ and $f\left( x\right) $ by their (usually,
maximal) values at the maximal value of $x=L/\Lambda $ in the calculation.
In other words, we will plot the normalized functions 
\begin{equation}
d\left( x\right) /d\left( x_{\max }\right) ,\ f\left( x\right) /f\left(
x_{\max }\right)  \label{g4}
\end{equation}%
where the values of the coefficients $d\left( x_{\max }\right) $ and $%
f\left( x_{\max }\right) $ are irrelevant for our purposes. In this way, all
the functions change from 0 at $x=0$ to 1 at $x=x_{\max }$ and can be
visually compared with each other. The values of $x_{\max }$ in our
computations vary from 35 to 100 which means that there are between 35 and
100 quantized minibands accessible to the particles.

The data below involve the Gaussian correlator $\left( \ref{a2}\right) $,
power-law correlator $\left( \text{\ref{ee2}}\right) $ with $\mu =1$ (the
Staras function), and the exponential correlator in real space $\left( \ref%
{e2}\right) $ (\textit{i.e., }the correlator\textit{\ }$\left( \text{\ref%
{ee3}}\right) $ with $\lambda =0.5$). We will refer to the latter two as the
power-law and exponential correlators without specifying $\mu $\ and $%
\lambda $.

\subsection{Mode coupling}

To underscore the effects of mode coupling, we start from the calculation
with the \textit{artificially} frozen mode coupling (the off-diagonal
components of the matrix of scattering probabilities $W_{jj^{\prime }}$ are
not calculated, but made equal to zero, $W_{jj^{\prime }}=W_{j}\delta
_{jj^{\prime }}$). This is a good starting point since in high quality films
with $R/L\gg 1$ the mode coupling is suppressed anyway. In this case the
transport equations $\left( \text{\ref{ee6}}\right) $ and $\left( \text{\ref%
{s2}}\right) $ can be solved analytically \cite{arm1}: 
\begin{equation}
\sigma \simeq \frac{2e^{2}}{3\hbar ^{2}m^{2}}\sum_{j}\frac{q_{j}^{2}}{%
W_{j}^{\left( 0\right) }-W_{j}^{\left( 1\right) }}  \label{ee10}
\end{equation}%
and similarly for diffusion, 
\begin{equation}
D\simeq \frac{1}{m^{3}S}\sum_{j}\frac{q_{j}^{2}}{W_{j}^{\left( 0\right)
}-W_{j}^{\left( 1\right) }},  \label{ee11}
\end{equation}%
where $W_{j}^{\left( 0,1\right) }$\ are the zeroth and first angular
harmonics of the transition probabilities $\left( \text{\ref{rrr3}}\right)
,\left( \text{\ref{t6}}\right) $. Note, that since $W_{j}$ for surface
correlators grows proportionally to $j^{4}$, Eq. $\left( \ref{rrr3}\right) $%
, the sum in Eq. $\left( \ref{ee10}\right) $ is rapidly convergent and the
contribution of the higher modes in systems without mode coupling can be
negligible. This means that for the surface scattering without mode coupling
the dependence of the conductivity $\sigma $ on the film thickness is
practically a smooth function, Figure 1 (small kinks on the curves cannot be
seen with the resolution of the figure). This is not so for the
single-particle diffusion $\left( \text{\ref{ee11}}\right) $ which contains
a step-wise factor $S\left( L\right) $, Figure 2. The small saw-tooth drops
on the curves $d\left( L\right) $ at the points in which $S=\mathrm{Int}%
\left( L/\Lambda \right) $ changes by 1 (quantum size effect, QSE) can serve
as useful markers on the curves that help to separate the standard QSE from
other phenomena. Therefore, more often than not we will plot $d\left(
L\right) $ rather than $\sigma \left( L\right) $. Figures 1 and 2 combine
the data for the Gaussian, exponential, and power-law inhomogeneities and
five different values of $R/L=0.1;1;10;50;100$. All normalized curves $%
f\left( x\right) /f\left( 50\right) $ and $d\left( x\right) /d\left(
50\right) $ are identical to each other except, of course, for the
normalization coefficients $f\left( 50\right) $ and $d\left( 50\right) $,
which change by the orders of magnitude depending on $R/L$ and the type of
inhomogeneities.

\subsection{Bulk \textit{vs.} surface scattering}

In Ref. \cite{pon1} we reported the existence of a new class of QSE for a
"boring" type of high-quality films $R\gg L$ with Gaussian or exponential
(in momentum space) surface and thickness fluctuations. This QSE manifests
itself as giant oscillations of $\sigma \left( L\right) $ with a relatively
large period which is directly related to the correlation radius of surface
inhomogeneities,%
\begin{equation}
L_{j}\simeq \frac{\pi }{2}\sqrt{\left( 2j+1\right) R\Lambda },  \label{eee11}
\end{equation}%
where $L_{j}$ are the positions of the peaks in $\sigma \left( L\right) $.

The effect is so large and well pronounced that a natural question is why
has not it been observed earlier in, for example, impurity scattering. As we
discussed in the Introduction and Section IV, the equations that describe
surface and bulk scattering are so similar that it looks like this new QSE
should be observed in bulk scattering as well. Surprisingly, this is not the
case. Figure 3 presents the normalized diffusion coefficient for surface and
bulk scattering. Both surface and bulk correlation functions are Gaussian
with the same large value of $R/L=50$. The curve with surface scattering
exhibits large QSE oscillations with peaks in positions $\left( \ref{eee11}%
\right) $, while the bulk scattering results in a monotonic dependence
similar to that for systems with frozen mode coupling in Figures 1 and 2.
This surprising difference between the bulk and boundary scattering requires
an explanation.

Our explanation of the "new" QSE, Eq. $\left( \ref{eee11}\right) ,$ in Ref. 
\cite{pon1} was that in high-quality films with $R/L\gg 1$ the mode coupling
is largely suppressed because the scattering-driven change in momentum $%
\delta q\sim 1/R$ is insufficient to induce transitions between the modes
which are widely separated between themselves at small $L$, $\delta \epsilon
\propto 1/L^{2}$. The mode coupling processes turn on one by one only at the
values of thickness given by Eq. $\left( \ref{eee11}\right) $. This should
be the same for both surface and bulk scattering. What is not the same is
the effect that this opening of the mode coupling channels has on transport.
In the case of surface scattering, the modes with the lowest quantum numbers
(grazing particles) make the largest contribution to transport (without the
quantum cut-off, the contribution from the particles in the lowest mode -
grazing particles moving parallel to the surface - would have been
infinite). Therefore, the threshold opening of individual mode coupling
channels for the lowest modes, which almost doubles the corresponding
cross-sections, is very noticeable in transport. In the case of bulk
scattering, the situation is different. Here all modes contribute more or
less equally to transport. Therefore, in the case of a large number of
available modes, the opening of few mode-coupling channels in points $\left( %
\ref{eee11}\right) $ is unnoticeable and the transport coefficients remain
nearly the same as in the pure diagonal case. This explains why this new QSE 
$\left( \ref{eee11}\right) $ has not been discovered earlier when studying
the bulk-dominated transport.

Figure 4 illustrates very different sensitivity of the surface- and
bulk-driven transport to mode coupling. The figure contains the normalized
single-particle diffusion coefficient in the cases of bulk and surface
scattering. In both cases the inhomogeneities are Gaussian with four
different values $R/L=0.1;1;10;50$. In the case of surface scattering, the
pattern evolves from the "usual" QSE saw-tooth curve for $R/L=0.1$ to the
new QSE with huge oscillations on more or less smooth curve for $R/L=50$
(the curves are marked by the value of $R/L$). In the case of bulk
scattering, all the curves split into two groups of smooth curves for small
and large $R/L$ (with, correspondingly, robust and mostly suppressed mode
coupling). Though the shapes of the curves from these two groups are
distinctly different, the difference in shapes, in contrast to surface
scattering, is rather quantitative than qualitative. Of course, the
saw-tooth anomalies, which are inherent to QSE, are more distinct on the
curves with robust mode coupling at small $R/L$.

Summarizing, QSE and the manifestations of the mode coupling are distinctly
different in surface- and bulk dominated quantized systems. This is due not
to the difference in mode coupling processes themselves, but due to the
different patterns for contributions from individual modes, especially for
the grazing particles. This also answers a puzzling question why the new
type of QSE is observed primarily in systems with the dominant surface
relaxation.

\subsection{Opening of mode coupling channels and mode contributions for
surface scattering}

The next step is the analysis of contributions from individual modes and
from the mode coupling transitions. We start from the data for the
artificially frozen mode coupling transitions in Figures 1,2 for surface
scattering and turn on such transitions. The results strongly depend on the
size of inhomogeneities $R/L$. Analysis of the scattering probabilities $W$
shows that for all reasonable types of inhomogeneities the decay of the
surface-driven scattering probabilities $W_{jj^{\prime }}$ at large $%
j,j^{\prime }$ is a function of $\left( j+j^{\prime }\right) ^{2}R^{2}/L^{2}$
and $\left( j-j^{\prime }\right) ^{2}R^{2}/L^{2}$. Therefore, for small
inhomogeneities, $R/L\ll 1$, the the mode-coupling scattering probabilities $%
W_{jj^{\prime }}$ with $j^{\prime }\neq j$ have the same order of magnitude
as for the intraband scattering $W_{jj}$. What is more, the contributions of
the higher modes to particle transport are quite noticeable and decrease
rather slowly with increasing $j,j^{\prime }$. This is illustrated in Figure
5 for the power-law inhomogeneities $\left( R/L=0.1\right) $ where three
curves correspond to diffusion in the single-mode, two-mode, and three-mode
regimes including the mode coupling. The surface correlator is exponential
in momentum space, Eq. $\left( \ref{ee2}\right) $, with $\mu =1$ (the Staras
correlator). All three curves are normalized by the single value of $d\left(
50\right) $ for the single-mode curve. As one can see from the Figure, the
turning on of the mode coupling does not lead to any qualitative changes and
results simply in the increase of the overall scattering cross-section. The
contributions from all modes have the same order of magnitude.

The situation changes dramatically at large $R/L$. Figure 6 presents the
same calculation as for Figure 5 but for $R/L=100$. As one can see, at small 
$x=L/\Lambda $ all three curves coincide. This means that the contribution
of higher modes and the mode-coupling effects are negligible even though the
interband transitions are allowed. This is explained by the fact that in
high-quality films with $R/L\gg 1$ the mode-coupling transitions switch on
one by one when the values of the film thickness reaches the values $\left( %
\ref{eee1}\right) $ (with a small logarithmic correction which depends on
the correlation function of inhomogeneities; Eq. $\left( \ref{eee1}\right) $
assumes that the quantum numbers $j$ are small and $q_{j}\sim \hbar /\Lambda 
$). Therefore, at thicknesses $L<L_{1}$\ (the first branching point in the
curves) all interband transitions $j\leftrightarrow j^{\prime }$ are
suppressed. Since at $R/L\gg 1$ the absolute values of $\left( jR/L\right)
^{2}$ grow very rapidly with $j$, the contributions of the higher modes are
negligible as well, and the system is effectively in a single-mode regime.
At $L\sim L_{1}$, Eq. $\left( \ref{eee1}\right) $, the mode-coupling
probability $W_{12}$ becomes comparable to $W_{11}$ and the mode-coupling
between the \emph{two} \emph{lowest} modes becomes robust. Only at this
point contribution of the second mode become noticeable. Therefore, at $%
L_{1}<L<L_{2}$ the system is effectively in a two-mode regime, at $%
L_{2}<L<L_{3}$ - in a three-mode regime, and so on, as it is seen clearly in
Figure 6.

The results for the Gaussian correlator are similar to those for the
power-law one. The exponential correlators, on the other hand, do not
exhibit behavior similar to Figure 6. The power spectrum for such
correlators is decaying very slowly at large $q$ thus ensuring robust
mode-coupling transitions at any $R/L$. As a result, the curves $d\left(
L\right) $ and $\sigma \left( L\right) $ resemble those on Figure 5 at any
value of $R/L$.

The important conclusion here is that in high-quality films $R/L\gg 1$ with
the surface-driven scattering the contribution of the higher modes becomes
important only as a result of the emergence of mode coupling at the values
of the film thickness $\left( \ref{eee1}\right) $. However, after the mode
coupling turns on at certain values of the film thickness, the contribution
of the higher mode becomes much bigger than it is usually believed.

\subsection{Particles with non-quadratic energy spectrum}

It is very interesting to check to what extent our analysis of the mode
coupling effects is sensitive to the form of particle spectra. We start from
deriving an analog of Eq. $\left( \ref{eee1}\right) $ for non-parabolic
particles, \emph{i.e}, from finding the critical values of thickness at
which the mode coupling become noticeable for individual modes in
high-quality films, $R/L\gg 1$.

Let us consider particles with a bulk spectrum $\epsilon \left( \mathbf{p}%
\right) $, or, after quantization, $\epsilon _{j}\left( \mathbf{q}\right)
=\epsilon \left( \pi \hbar j/L,\mathbf{q}\right) $. Scattering by
inhomogeneities of the lateral size $R$ change the lateral momentum by $%
\delta q\sim \hbar /R$. This small change in momentum ($R\gg L$) is
sufficient for the interband transition $j\leftrightarrow j+1$ only if the
energy conservation, $\epsilon _{j}\left( \mathbf{q}\right) =\epsilon
_{j+1}\left( \mathbf{q-\hbar /}R\right) $, can be satisfied:%
\begin{equation}
0=\frac{\partial \epsilon _{j}}{\partial j}-\frac{\hbar }{R}\frac{\partial
\epsilon _{j}}{\partial q}.  \label{g6}
\end{equation}%
The solution of this equation gives the values of the critical thickness $%
L_{j}$ at which the mode coupling channels $j\leftrightarrow j+1$ become
open. [More accurate equation than $\left( \ref{g6}\right) $\ contains
coefficients of the order of 1 which depend on detail of the correlation
function of inhomogeneities]. In the case of parabolic inhomogeneities, as
for all particles for which the energy spectrum\textit{\ }$\epsilon \left( 
\mathbf{p}\right) $ depends only on the absolute value of momentum $p$, Eq. $%
\left( \ref{g6}\right) $\ reduces to $\left( \ref{eee1}\right) $.

As an example, we consider "ultrarelativistic" particles, $\epsilon =cp$, 
\textit{i.e.,}%
\begin{equation}
\epsilon _{j}\left( \mathbf{q}\right) =c\sqrt{\left( \pi \hbar j/L\right)
^{2}+q^{2}}.  \label{g7}
\end{equation}%
The applications include photons between two (rough) mirrors or quantized
phonons in helium films. For non-parabolic spectra $\epsilon \left( p\right) 
$, the equation $\left( \ref{rrr3}\right) $ for the scattering probabilities 
$W_{jj^{\prime }}$ should be modified as \cite{arm2}%
\begin{equation}
W_{jj^{\prime }}\left( \mathbf{q},\mathbf{q}^{\prime }\right) \delta \left(
\epsilon _{j\mathbf{q}}-\epsilon _{j^{\prime }\mathbf{q}^{\prime }}\right) =%
\frac{j^{2}j^{\prime 2}}{L^{2}\hbar ^{3}}\frac{\left( \epsilon _{j^{\prime }%
\mathbf{q}}-\epsilon _{j\mathbf{q}^{\prime }}\right) ^{2}}{\left(
j^{2}-j^{\prime }{}^{2}\right) ^{2}}\left[ \zeta _{11}+\zeta _{22}+2\zeta
_{12}\left( -1\right) ^{j+j^{\prime }}\right] \delta \left( \epsilon _{j%
\mathbf{q}}-\epsilon _{j^{\prime }\mathbf{q}^{\prime }}\right) ,  \label{g9}
\end{equation}%
while $q_{j}/m$ in the equations for the single-particle diffusion $\left( %
\ref{s2}\right) $ should be replaced by the lateral velocity $v_{j}=\partial
\epsilon _{j}/\partial q$. Then straightforward algebra leads to the
following expression to the single-particle diffusion coefficient:%
\begin{eqnarray}
D &=&\frac{cR^{3}}{\ell ^{2}}d\left( \omega L/\pi c\right) ,  \label{g10} \\
d\left( x\right) &=&\frac{1}{2\pi ^{6}x^{3}S\left( x\right) }\left( \frac{R}{%
L}\right) ^{3}\left( \frac{q_{j}R}{\hbar }\right) \widetilde{\tau
_{jj^{\prime }}}^{-1}\left( \frac{q_{j^{\prime }}R}{\hbar }\right)  \notag
\end{eqnarray}%
where $\widetilde{\tau _{jj^{\prime }}}^{-1}$ is the dimensionless inverse
relaxation time matrix $\left( \ref{ee6}\right) $ with the changes mentioned
above.

Figure 7 presents normalized diffusion coefficient $D\left( x\right)
/D\left( 50\right) $, Eq. $\left( \ref{g10}\right) $, for random
inhomogeneities with the Gaussian correlation function; $x=\omega L/\pi c$
is the dimensionless energy (frequency). The curves are marked by the values
of $R/L$. The curves with small $R/L$ exhibit the "standard" saw-tooth QSE.
The curves for larger inhomogeneities exhibit the large-scale oscillations
with the peaks which correspond to opening of the mode-coupling channels and
which are described approximately by Eq. $\left( \ref{g6}\right) $. The
curves for the exponential in momentum space surface correlators $\left( \ref%
{ee2}\right) $ are, essentially, the same. If the surface correlator has a
power-law shape in momentum space, which means that there exist
inhomogeneities of all sizes, the transport coefficients for particles with
non-quadratic spectra assume the same "standard" saw-tooth shape as for the
particles with a parabolic spectrum.

The general conclusion is that the mode coupling effects, which manifest
themselves in the new type of QSE in transport in high-quality films, are
very robust and are not sensitive to the type of the energy spectrum. This
means the this type of QSE should exist for (quasi-)particles of different
nature and for various solid state systems.

\subsection{Interwall interference effects}

What also makes the scattering by surface inhomogeneities different from
scattering by bulk fluctuations or impurities is the possible interference
of particle reflected from the opposite wall. This interference is
especially interesting in the case when the inhomogeneities from the
opposite walls are correlated. In this Section we discuss the effect of this
interwall correlation of inhomogeneities on mode coupling. The existence of
this non-trivial effect is a unique feature of surface scattering that does
not have any analog in scattering by bulk inhomogeneities. Surprisingly, the
possibility of cross-wall correlation of surface inhomogeneities from
opposite walls gives an interesting insight into mode coupling. The study of
the effect of interwall correlation of inhomogeneities has been initiated in
Refs.\cite{arm1,arm3,pon1} (for additional results in application to
excitons see Ref. \cite{lis1}).

The effect of interwall correlations has two unique features stemming from
the sign of the interference of scattering from opposite walls. Because of
the $\left( -1\right) ^{j+j^{\prime }}$ factor in the interwall contribution
to the scattering probability, Eq.$\left( \ref{rrr3}\right) $, the
contribution of the interwall correlation function $\zeta _{12}$ has
different signs for in-band $\left( j=j^{\prime }\right) $ and mode-coupling 
$\left( j=j^{\prime }\pm 1\right) $\ scattering processes. Depending on the
magnitude of $\zeta _{12}$ and \emph{its sign}, its contribution can enhance
or suppress the mode coupling effects.

To decrease the number of parameters, we assume that, as in Refs.\cite%
{arm1,arm3,pon1}, the correlation functions of inhomogeneities on both walls 
$\zeta _{11}$ and $\zeta _{22}$ are given by the same function, $\zeta
_{11}\left( s\right) =\zeta _{22}\left( s\right) =\zeta \left( s\right) $.
The structure of the interwall correlator of inhomogeneities $\zeta
_{12}\left( s\right) $ is assumed to be the same as for the intrawall
correlations with the same correlation radius $R$. However, the amplitude $a$
of the interwall correlations is different from the intrawall ones, 
\begin{equation}
\zeta _{11}=\zeta _{22}=\zeta \left( s\right) ,\ \zeta _{12}\left( s\right)
=a\zeta \left( s\right) ,\ \left\vert a\right\vert \leq 1.  \label{ii1}
\end{equation}%
Note that in contrast to the on-wall correlation functions $\zeta
_{11},\zeta _{22}$, the sign of the interwall correlation function $\zeta
_{12}$ is not fixed; even $\zeta _{12}\left( s=0\right) $ can be negative.
By itself, the sign of the interwall correlations $\zeta _{12}=\left\langle
\xi _{1}\cdot \xi _{2}\right\rangle $ is ambiguous and depends on how do we
introduce the signs of the deviations of the wall positions $\xi _{1,2}$
from the averages $\pm L/2$; throughout this paper, we use the definitions $%
\left( \ref{a0}\right) $. With this definition, the sign of $a$ can be
positive or negative depending on whether the inhomogeneities from the
opposite walls "attract" or "repel" each other. If the inhomogeneities from
the opposite walls simply reproduce each other ("parallel" walls; the film
thickness is constant along the film), then, with our definition of the wall
inhomogeneities Eq.$\left( \ref{a0}\right) $, $a=-1$. This type of interwall
correlation is likely to occur when an ultrathin film grows on an
inhomogeneous substrate. In the opposite case of walls with opposite
modulations ("antiparallel" walls), $a=1$. This is the case of pure
thickness fluctuations, which is likely to occur, for example, after the
film (wire) has been inhomogeneously stretched. In the case of "parallel"
walls, $a\rightarrow -1$, the destructive interference of scattering by
opposite walls can, in the absence of mode coupling (see below), completely
negate all transport manifestations of the wall corrugation. These two
limiting situations is presented in Figure 8. In general, $-1\leq a\leq 1$.

To extract the effect of interwall correlations, we will present the results
for the transport coefficient at different values of the interwall amplitude 
$a$ and calculate the relative change of conductivity $\sigma $ and
diffusion coefficient $D$ caused by\ the introduction of such correlations, 
\begin{equation}
\phi ^{\left( a\right) }=\frac{f^{\left( a\right) }-f^{\left( 0\right) }}{%
f^{\left( 0\right) }},\frac{d^{\left( a\right) }-d^{\left( 0\right) }}{%
d^{\left( 0\right) }},  \label{i1}
\end{equation}%
where $f^{\left( a\right) }$ and $f^{\left( 0\right) }$ ($d^{\left( a\right)
}$ and $d^{\left( 0\right) }$)\ are the values of $\sigma $ and $D,$ Eqs. $%
\left( \ref{g1}\right) ,\left( \ref{g2}\right) $,\ calculated with and
without interwall correlations. An additional benefit of this definition is
that the functions $\phi ^{\left( a\right) }$ are automatically normalized.

In the presence of interwall correlations $\left( \ref{ii1}\right) $, the
transition probabilities $W_{jj^{\prime }}\left( \mathbf{q},\mathbf{q}%
^{\prime }\right) $ $\left( \text{\ref{rrr3}}\right) $\ become proportional
to 
\begin{equation}
2\left[ 1+a\left( -1\right) ^{j+j^{\prime }}\right] \zeta \left( \left\vert 
\mathbf{q}_{j}-\mathbf{q}_{j^{\prime }}^{\prime }\right\vert \right) .
\label{i2}
\end{equation}%
The most interesting effects of the interwall correlations are related to
the oscillating structure of the term with $a$ in Eq. $\left( \text{\ref{i2}}%
\right) $.

If the mode coupling is suppressed, then the only important terms in Eq. $%
\left( \text{\ref{i2}}\right) $ are the diagonal ones with $j=j^{\prime }$
and the function $\phi ^{\left( a\right) }$ is a constant,%
\begin{equation}
\phi ^{\left( a\right) }\left[ \text{\textit{no mode coupling}}\right] =%
\frac{1}{1+a}-1=\frac{-a}{1+a}.  \label{i3}
\end{equation}%
In this case the presence of interwall correlations leads to a simple
increase or decrease, depending on the sign of $a$, of the transport
coefficients by the factor $-a/(1+a)$. This is always the case, for example,
when only one mode is important \cite{lis1}. Therefore, all deviations of $%
\phi ^{\left( a\right) }$\ from the constant $\left( \text{\ref{i2}}\right) $%
\ are due solely to the scattering-driven mode coupling. This gives a
non-trivial insight into the mode coupling and its consequences.

For example, as it is clear from Figure 6, the mode coupling for the
power-law inhomogeneities with $R/L=100$ appears only at $x=L/\Lambda >15$.
Therefore, the function $\phi ^{\left( a\right) }\left( x<15\right) $ should
be flat, Eq. $\left( \text{\ref{i3}}\right) $, and exhibit anomalies in the
points in which the mode coupling effects are switched on. This is
illustrated in Figure 9\textbf{\ }which presents the function $\phi ^{\left(
a\right) }\left( x\right) $ for the same power-law inhomogeneities as Figure
4 ($R/L=100$) at five different values of $a$, $a=-0.5;-0.1;0.1;0.5;09.$ The
flat parts on left hand side of all curves $\left( x<15\right) $ correspond
to the absence of mode coupling at these values of $R/L$ and are given by
Eq. $\left( \text{\ref{i3}}\right) $. The peaks in the curves show the
values of $x=L/\Lambda $ for the consecutive openings of the mode scattering
channels. The difference in the amplitudes of the peaks is easily explained
by the dependence of the scattering probabilities on the interwall
correlation amplitude $a$, Eq. $\left( \text{\ref{i2}}\right) $.

The contribution of the term with $a$\ in $\left( \text{\ref{i2}}\right) $
has a different sign for different $W_{jj^{\prime }}$ depending on whether $%
j+j^{\prime }$\ is even or odd. Since the mode coupling channels $%
j\longleftrightarrow j+1$ in high-quality films with Gaussian and
exponential power functions turn on one by one with increasing $L$, one
would expect that the function $\phi ^{\left( a\right) }\left( L\right) $\
in such films should become a step-wise function of $L$. This is not
correct. In films with large $R/L$ and frozen out mode coupling effects the
contributions of individual modes decrease rapidly as $1/j^{4}$ \cite{arm1}.
However, when the transitions $j\longleftrightarrow j+1$ are switched on,
the overall contribution of the mode $j+1$ increases disproportionately
(Figure 6). As a result, the function $\phi ^{\left( a\right) }$ becomes an
oscillating rather than step-wise function as it is seen clearly in Figure 9.

The positions of the QSE peaks in systems with interwall correlations differ
from Eq. $\left( \text{\ref{eee11}}\right) $ and depend on the value of $a$.
The shifts of peaks in Figure 9, which depend on the value of $a$, are
better illustrated in Figure 10 in which we presented the normalized
diffusion coefficient itself (and not the function $\phi ^{\left( a\right) }$%
) for the same type of surface inhomogeneities and the same value of $%
R/L=100 $ for two different interwall amplitudes, $a=-0.9;0.9$ as a function
of $x=L/\Lambda $. The explanation of these shifts is the following. Let us
assume that the first peak is observed at $x=x_{1}$. In this point the value
of $W_{12}\left( x\right) $ reaches $W_{11}\left( x\right) $, $W_{11}\left(
x_{1}\right) =W_{12}\left( x_{1}\right) .$ According to Eq. $\left( \text{%
\ref{i2}}\right) ,$ in the presence of interwall correlations, these
scattering probabilities $W^{\left( a\right) }$ change with respect to their
values $W^{\left( 0\right) }$ in the absence of interwall correlations as%
\begin{equation}
W_{11}^{\left( a\right) }=\left( 1+a\right) W_{11}^{\left( 0\right) },\
W_{12}^{\left( a\right) }=\left( 1-a\right) W_{12}^{\left( 0\right) }.
\label{i4}
\end{equation}%
Since near the peak position $W_{12}^{\left( 0\right) }\left( x\right) $
grows very rapidly while $W_{11}^{\left( 0\right) }$ does not change much, $%
W_{12}^{\left( a\right) }\left( x\right) $ reaches the value $W_{11}^{\left(
a\right) }$ earlier than $W_{12}^{\left( 0\right) }\left( x\right) $ reaches 
$W_{11}^{\left( 0\right) }$\ at negative $a$ and later at positive $a$. This
exactly what is happening in Figure 10. At small $a$, the change in position 
$x_{1}$ with respect to its value in the absence of interwall correlations is%
\begin{equation}
\Delta x_{1}=2aW_{11}^{\left( 0\right) }\left( x_{1}\right) /\left[ \partial
W_{12}^{\left( 0\right) }/\partial x-\partial W_{11}^{\left( 0\right)
}/\partial x\right] .  \label{i5}
\end{equation}

The oscillating nature of the interwall contribution, Eq. $\left( \text{\ref%
{i2}}\right) $ should be more pronounced for the systems with smaller
inhomogeneities, in which the mode-coupling transitions are as probable as
the intraband scattering. In this case the flat areas $\left( \text{\ref{i3}}%
\right) $ should be absent. Instead, the curves $\phi ^{\left( a\right)
}\left( x\right) $ should exhibit QSE with the oscillations in points $x_{S}$
in which the number of occupied minibands $S$ changes by 1, $S\rightarrow
S+1 $. [With our choice of dimensionless variables, the period of these
oscillations is equal to 1]. This is illustrated in Figures 11 and 12 which
contain the data similar to those in Figure 9 but for smaller
inhomogeneities, $R/L=0.1$ and $R/L=1$ respectively. At $R/L\ll 1$, when all
mode coupling transitions and intraband scattering are equally probable, the
addition of an extra band $S$ adds all interwall terms with the
sign-changing coefficients $a\left( -1\right) ^{j+S}$, Eq. $\left( \text{\ref%
{i2}}\right) $. However, since the main mode, $j=1$, contributes the most to
the transport flow, the overall sign of the interwall contribution is the
sign of $a\left( -1\right) ^{1+S}$ and should change in the points in which $%
S\left( x\right) $ changes. This is exactly what can be seen in Figure 11.
The amplitude of the oscillations grows with an increase in $\left\vert
a\right\vert $ and goes down with increasing $x$. Figure 11\textbf{\ }($%
R/L=0.1$)\textbf{\ }demonstrates these oscillations for large interwall
correlations, $a=-0.9;0.9$. Since the signs of these two interwall
amplitudes are opposite, the contributions from these two types of
cross-correlations have opposite signs, Eq. $\left( \text{\ref{i2}}\right) $%
. The analogous curves for all interwall amplitudes $\left\vert a\right\vert
<0.9$ are squeezed between the curves for $a=-0.9;0.9$. At larger
inhomogeneities, $R/L=1$ (Figure 12), one can still see the well-pronounced
QSE oscillations, but the average is already noticeably shifted from zero as
it should be at larger $R/L$ (\textit{cf. }Figure 9).

Note, that the height of the first peak is always given by Eq. $\left( \text{%
\ref{i3}}\right) $ and can be quite large when $a\rightarrow -1$ ("parallel"
walls). At $a\rightarrow -1$ the interwall correlation compensates almost
completely for dephasing caused by scattering from individual wall
inhomogeneities. In this case, if the wall scattering is the only relaxation
mechanism (ballistic transport), the lateral mean free path becomes
infinitely large \emph{even if both walls are rough!} In Figures 11 and 12
the height of the first peak for $a=-0.9$ is 9 and the peak does not fit
into the Figures. For positive values of the interwall amplitude $a$
("antiparallel" walls), the second peak has the largest amplitude, while the
first (negative) one, which is given by Eq. $\left( \text{\ref{i3}}\right) $%
, has a smaller amplitude.

In conclusion, the possible correlation of random inhomogeneities from the
opposite walls provides a non-trivial insight into the mode coupling. On the
other hand, measurements of the dependence of the transport coefficients on
the film thickness or particle energy can provide unique information on the
interwall correlations since, depending on the situation, the effect of
interwall correlations can be constructive, destructive, or oscillating. The
shift of oscillations gives the information on both the strength and sign of
the interwall correlations.

\subsection{Scattering by multiscale inhomogeneities}

Above we studied the systems with random inhomogeneities of a single,
well-defined spatial scale (correlation radius) $R$. In the case of
single-scale inhomogeneities, such as inhomogeneities with a Gaussian or
exponential power spectrum, the mode coupling channels at large $R$ open one
by one at definite values of the film thickness $\left( \text{\ref{eee11}}%
\right) $ leading to large scale QSE oscillations of the transport
coefficients. In the opposite case of the power spectrum with
inhomogeneities of all sizes, such as slowly decaying power-law power
spectrum with a low index, the mode coupling is always robust leading to the
disappearance of the large-scale oscillations and the restoration of the
standard saw-tooth QSE. It is interesting to investigate the behavior of QSE
in an intermediate situation in system with inhomogeneities of few distinct
scales. Figure 13 presents the data for the diffusion coefficient for a film
with Gaussian surface inhomogeneities of three types: the inhomogeneities
with $R/L=25$, $R/L=10$, and the inhomogeneities with a combination of both
sizes (the sum of the corresponding power spectra). All three curves are
normalized by the same value $d\left( R/L=25;x=50\right) $ and are labeled
by the value of $R/L$. As one can easily see, the combining the
inhomogeneities with two correlation sizes does not lead to a mechanical
mixture of the individual oscillations but results in smoothing, shifting,
and rescaling of the oscillations. Adding several more scales leads simply
to a disappearance of the QSE oscillations. At present it is not clear yet
the combination of how many scales is necessary for the restoration of the
saw-tooth behavior.

\section{Conclusions}

In summary, we analyzed the mode coupling and its effect on QSE in transport
in high quality quasi-$2D$ quantum systems ($R/L\gg 1$) with various types
of surface, thickness, and bulk fluctuations. Here are the conclusions:

\begin{itemize}
\item Though the transport equations and mode coupling effects for systems
with bulk and interface fluctuations look almost identical, QSE in such
systems is not the same. The appearance of large scale oscillations of the
transport coefficients requires not only the opening of mode coupling
channels at distinct values of film thickness $\left( \text{\ref{eee11}}%
\right) $, $\left( \text{\ref{g6}}\right) $ as it happens in both surface-
and bulk-driven systems, but also the predominant role of modes with low
quantum numbers (grazing particles). The latter requirement is routine for
surface scattering but is not fulfilled for bulk fluctuations of a general
form. Only if the fluctuations in the bulk do not depend on the coordinate
across the film, Eq. $\left( \text{\ref{t7}}\right) $, the transport
coefficients manifest the same large-scale oscillations as in the case of
surface scattering. This explains a huge difference in QSE in high-quality
films with bulk and surface scattering.

\item One of the most striking conclusions concerns the contributions from
different modes in high quality samples in the case of surface scattering by
large inhomogeneities, $R/L\gg 1$. Without mode coupling, the contributions
from a mode with quantum number $j$ would be proportional \cite{arm1} to $%
1/j^{4}$ and higher modes would have been almost irrelevant. These higher
modes contribute to transport \emph{only} because of the mode coupling. This
conclusion is of little interest for small-size defects $R/L\ll 1$ since in
such systems the scattering-driven mode coupling is always robust and the
contribution from the higher modes is important.

In high-quality films $R/L\gg 1$, the mode-coupling transitions switch on
one by one at the values of the film thickness $\left( \text{\ref{eee11}}%
\right) $, $\left( \text{\ref{g6}}\right) $. Thus, the higher modes become
important also one by one, only after the corresponding mode coupling
channel is turned on, Figure 6. After the mode coupling turns on, the
contribution of the higher modes is much higher than one usually assumes and
the description that singles out the grazing particles becomes wrong.

\item The consecutive opening of the mode coupling channels in high-quality
films at distinct values of the film thickness, Eqs. $\left( \text{\ref%
{eee11}}\right) $, $\left( \text{\ref{g6}}\right) $, which leads to giant
QSE oscillations, is a very robust effect that is not very sensitive to the
nature of (quasi-)particles and the form of their spectrum. As a result, the
effect can be observed in a wide variety of quantized systems such as metal
or semiconductor films, quantum wires, ultranarrow channels, optical fibers, 
\emph{etc}.

\item An interesting manifestation of the coupling effects in high quality
films is related to possible correlation between random inhomogeneities from
the opposite walls. The interference of scattering from the opposite walls
changes its sign from constructive to destructive depending on the parity of
the sum of mode quantum numbers and, therefore, provides contributions of
the opposite signs for intramode and mode coupling channels. In some cases
("parallel" walls) the opening of the mode coupling channel can be
responsible tor the cutoff for the mean free path for grazing particles
which would be nearly divergent otherwise. The presence of interwall
correlations can help to distinguish films with surface and thickness
fluctuations. The shift of conductivity or diffusion oscillations provides
the information on both the sign and strength of interwall correlations.

\item The presence of multiscale inhomogeneities with several distinct
correlation radii $R$ leads, instead of a mechanical mixture of individual
QSE\ patterns, to shifting and smoothing of the QSE oscillations of the
transport coefficients that are inherent to high-quality films with a
single-scale roughness.

\item The results can lead to new, non-destructive ways of studying the
quality of the high quality surfaces, including the buried surfaces and
interfaces, by measuring the lateral conductivity or diffusion. This is
especially valuable for high-quality surfaces with large-scale
inhomogeneities for which the usual scanning techniques can become
problematic because of very large scanning areas.
\end{itemize}

The results of the paper can be applied to particles in a wide range of
quantum quasi-$2D$ systems. The results can also be cautiously extended even
to quasi-$1D$ systems up to the point when the strong localization effects
render transport calculations meaningless (see review \cite{izrailev1} and
references therein).

\section{Acknowledgments}

The work was supported by NSF grant DMR-0077266.

\newpage

\section{Figure captions}

Figure 1. (Color online) Normalized conductivity $f\left( x\right) /f\left(
50\right) ,$ $x=L/\Lambda _{F}$, Eq. $\left( \ref{g1}\right) $,with
artificially frozen mode coupling. The shapes of all five curves are
identical irrespective of the type of inhomogeneities (Gaussian,
exponential, or power-law) and the value of $R/L$. With this resolution all
the curves are smooth.

Figure 2. (Color online) Normalized single-particle diffusion coefficient $%
d\left( x\right) /d\left( 50\right) ,$ $x=L/\Lambda $ with artificially
frozen mode coupling. The shapes of all five curves are identical
irrespective of the type of inhomogeneities (Gaussian, exponential, or
power-law) and the value of $R/L$. Small saw-tooth anomalies correspond to
changes in the number of accessible mini-bands $S\left( L\right) $, Eq. $%
\left( \ref{g3}\right) $.

Figure 3. (Color online) Normalized single-particle diffusion coefficient
with surface and bulk scattering; in both cases the inhomogeneities are
Gaussian with $R/L=50$.

Figure 4. (Color online) Normalized single-particle diffusion coefficient
with surface and bulk scattering; in both cases the inhomogeneities are
Gaussian with $R/L=0.1;1;10;50$. For surface scattering, all four curves,
marked by the values of $R/L$, are different. For bulk scattering, there are
two groups of coinciding curves with small and large $R/L$.

Figure 5. (Color online) Normalized single-particle diffusion coefficient $%
d\left( x\right) /d\left( 36\right) ,$ $x=L/\Lambda $ for power-law
inhomogeneities, Eq. $\left( \text{\ref{ee2}}\right) $ at $\mu =1$ for small
size inhomogeneities, $R/L=0.1$. Curve 1 takes into account only the main
mode. Curve 2 accounts for the first two modes, including coupling, curve 3
- the first three modes. It is clear that all three modes are equally
important. All three curves use \textit{the same }normalization parameter $%
d\left( 36\right) $ taken from the single (main) mode contribution (curve 1).

Figure 6. (Color online) The same as Figure 5 but for large-scale
inhomogeneities, $R/L=100$. The splits occur in the points when the mode
coupling becomes noticeable. It is clear that the contributions of the
higher modes become noticeable only when their coupling to the main mode
becomes large.

Figure 7. (Color online) Normalized diffusion coefficient $D\left( x\right)
/D\left( 50\right) $, Eq. $\left( \ref{g10}\right) $, for ultrarelativistic
particles, $\epsilon =cp$, and random inhomogeneities with the Gaussian
correlation function. The curves are marked by the values of $R/L$; $%
x=\omega L/\pi c$ is the dimensionless energy (frequency).

Figure 8. (Color online) Two different types of walls with correlated random
inhomogeneities. For "parallel" walls the interwall correlation amplitude $%
a=-1$, for "antiparallel" walls - $a=1$. In general, the interwall
correlation amplitude $-1\leq a\leq 1$.

Figure 9. (Color online) Relative contribution of the interwall correlations
to the single-particle diffusion, Eq. $\left( \text{\ref{i1}}\right) $.
Large-size ($R/L=100$) power-law inhomogeneities, Eq. $\left( \text{\ref{ee2}%
}\right) $ with $\mu =1$. The curves are labelled by the values of the
interwall amplitude $a$.

Figure 10. (Color online) Normalized single-particle diffusion coefficient $%
d\left( x\right) /d\left( 36\right) $ for the same inhomogeneities as in
Figure 8 for two values of the interwall amplitude $a$, $a=-0.9;0.9$.

Figure 11. (Color online) The same as in Figure 8\textbf{\ }but for small
inhomogeneities, $R/L=0.1.$ The curves are labelled by the values of the
interwall amplitude $a=-0.9;0.9$.

Figure 12. (Color online) The same as in Figure 8\textbf{\ }but with $R/L=1.$
The curves are labelled by the values of the interwall amplitude $a$.

Figure 13. (Color online) Diffusion coefficient for Gaussian inhomogeneities
with $R/L=10,$ $R/L=25$ and with the sum of inhomogeneities of both sizes.
The curves are labled by the value of $R/L$. All three curves are normalized
by the value of $d\left( R/L=25;x=50\right) .$


\begin{thebibliography}{99}
\bibitem{and1} P. A. Lee and T. V. Ramakrishnam, Rev.Mod.Phys. \textbf{57}%
,287 (1985); B. L. Altshuler, P. A. Lee, and R. A. Webb, (North-Holland,
Amsterdam, 1991); \textit{Scattering and Localization of Classical Waves in
Random Media, }ed. P. Sheng (World Scientific, Singapore, 1990)

\bibitem{matr1} K. A. Muttalib, J. -L. Pichard, and A. D. Stone,
Phys.Rev.Lett. 59, 2475 (1985); C. W. J. Beenakker, Rev.Mod.Phys. \textbf{69}%
, 731 (1997)

\bibitem{mar1} J. A. Sanchez-Gil, V. Freilikher, I. Yurkevich, and A. A.
Maradudin, Phys.Rev.Lett., \textbf{80}, 948 (1998)

\bibitem{arm2} A. E. Meyerovich, and A. Stepaniants, Phys. Rev. B \textbf{60}%
, 9129 (1999)

\bibitem{pon1} A. E. Meyerovich, and I. V. Ponomarev, Phys. Rev. B \textbf{65%
}, 155413 (2002)

\bibitem{q2} J.A.Ogilvy, \textit{Theory of Wave Scattering from Random
Surfaces} (Adam Hilger, Bristol) 1991

\bibitem{fer1} R. M. Feenstra, D. A. Collins, D. Z. Y. Ting, M. W. Wang, and
T. C. McGill, Phys. Rev. Lett. \textbf{72}, 2749 (1994)

\bibitem{fish2} G.Fishman, and D.Calecki, Phys. Rev. B \textbf{43}, 11 581
(1991)

\bibitem{pal1} G. Palasantzas, J. Barnas, Phys. Rev. B \textbf{56}, 7726
(1997); G. Palasantzas, Y.-P. Zhao, G.-C. Wang, T.-M. Lu, J. Barnas, and J.
Th. M. De Hosson, Phys. Rev. B \textbf{61}, 11 109 (2000)

\bibitem{conduct1} M. Jalochowski, M. Hoffmann, and E. Bauer, Phys.Rev.Lett.%
\textbf{\ 76}, 4227 (1996); Phys. Rev. B \textbf{51}, 7231 (1995); L. A.
Kuzik, Yu. E. Petrov, F. A. Pudonin, and V. A. Yakovlev, Sov.Phys. - JETP 
\textbf{78}, 114 (1994); G. M. Mikhailov, I. V. Malikov, and A. V. Chernykh,
JETP Lett. \textbf{66}, 725 (1997)

\bibitem{spect1} J. J. Paggel, T. Miller, and T. C. Chang, Science, \textbf{%
283}, 1709 (1999); D. A. Evans, M. Alonso, R. Cimino, and K. Horn,
Phys.Rev.Lett. \textbf{70}, 3483 (1993); J. E. Ortega, F. J. Himpsel, G. J.
Mankey, and R. F. Willis, Phys.Rev. B\textbf{47}, 1540 (1993)

\bibitem{susc1} S. Andrieu, C. Chatelain, M. Lemine, B. Berche, and Ph.
Bauer, Phys.Rev.Lett. \textbf{86}, 3883 (2001)

\bibitem{stm1} I. B. Altfeder, D. M. Chen, and K. A. Matveev, Phys.Rev.Lett. 
\textbf{80}, 4895 (1998); I. B. Altfeder, K. A. Matveev, and D. M. Chen,
Phys.Rev.Lett. \textbf{78}, 2815 (1997)

\bibitem{rr6} V.B.Sandomirskii, Sov. Phys.-JETP \textbf{25}, 101\ (1967)
[Zh. Eksp. \&Teor. Fiz. \textbf{52}, 158 (1968)]

\bibitem{r4} N.Trivedi, and N.W.Ashcroft, Phys. Rev. B \textbf{38}, 12298
(1988)

\bibitem{qse2} M. Jalochowski, E. Bauer, H. Knoppe, and G. Lilienkamp,
Phys.Rev. B \textbf{45}, 13607 (1992); M. Jalochowski, M. Hoffmann, and E.
Bauer, Phys.Rev. B\textbf{\ 51}, 7231 (1995); H. Sakaki, T. Noda, K.
Hirakawa, M. Tanaka, and T. Matsusue, Appl. Phys. Lett. \textbf{51}, 1934
(1987); L.-W. Tu, G. K. Wong, and J. B. Ketterson, Appl. Phys. Lett. \textbf{%
55}, 1327 (1989)

\bibitem{gu1} S. L. Phillipson, A. M. Guenault, S. N. Fisher, G. R. Pickett,
and P. J. Y. Thibault, \textit{Nature} \textbf{395}, 578 (1998) and P. A.
Reeves, A. M. Guenault, S. N. Fisher, G. R. Pickett, and G. Tvalashvili, 
\textit{Physica B} \textbf{284-288} 319 (2000)

\bibitem{reyer1} A. E. Meyerovich, and R. Jochemsen, J. Low Temp. Phys., 
\textbf{126}, 193 (2001)

\bibitem{neutron1} V.V.Nesvizhevsky, H.G.B\"{o}rner, A.K.Petukhov, H.Abele,
S.Bae\ss ler, F.J.Rue\ss , Th.St\"{o}ferle, A.Westphal, A.M.Gagarsky,
G.A.Petrov, and A.V.Strelkov\textit{,} Nature, \textbf{415}, 297 (2002);
V.V.Nesvizhevsky, H.G.B\"{o}rner, A.M.Garganski, A.K.Petoukhov, G.A.Petrov,
H.Abele, S.Bae\ss ler, G.Divkovic, F.J.Rue\ss , T.St\"{o}ferle, A.Westphal,
A.V.Strelkov, K.V.Protasov, A.Yu.VoroninPhys. Rev. D \textbf{67}, 102002
(2003)

\bibitem{arm1} A. E. Meyerovich, and A. Stepaniants, Phys. Rev. B \textbf{58}%
, 13 242 (1998)

\bibitem{arm3} A. E. Meyerovich, and A. Stepaniants, J. Phys.: Cond.Matt. 
\textbf{12}, 5575 (2000)

\bibitem{r3} Z.Tesanovic, M.V.Jaric, and S.Maekawa, Phys. Rev. Lett. \textbf{%
57}, 2760 (1986)

\bibitem{qq17} G.Fishman, and D.Calecki, Phys. Rev. Lett., \textbf{62}, 1302
(1989)

\bibitem{kaw1} A.Kawabata, J. Phys. Soc. Jap., \textbf{62}, 3988 (1993)

\bibitem{r2} A.E.Meyerovich, and S.Stepaniants, Phys. Rev. Lett. \textbf{73}%
, 316 (1994); Phys. Rev. B \textbf{51}, 17116 (1995); J. Phys.: Cond. Matt. 
\textbf{9}, 4157 (1997)

\bibitem{mak1} N.M.Makarov, A.V.Moroz, and V.A.Yampol'skii, Phys. Rev. B 
\textbf{52}, 6087 (1995)

\bibitem{arm4} A. E. Meyerovich, and A. Stepaniants, Aust. J. \ Phys., 
\textbf{53}, 53 (2000)

\bibitem{or1} J. Henz, H. von K\"{a}nel, M. Ospelt, and P. Wachter, Surf.
Sci. \textbf{189/190}, 1055(1987); J. Y. Duboz, P. A. Badoz, E. Rosencher,
J. Henz, M. Ospelt, H. von K\"{a}nel,, and A. Briggs, Appl. Phys. Lett. 
\textbf{53}, 788

\bibitem{coupl1} J. J. Paggel, T. Miller, and T. C. Chiang, Phys.Rev.Lett. 
\textbf{81}, 5632 (1998); F. Patthey and W.-D. Schneider, Phys.Rev. B 
\textbf{50}, 17560 (1994); M. Schmid, W. Hebenstreit, P. Varga, and S.
Crampin, Phys. Rev. Lett. \textbf{76}, 2298 (1996)

\bibitem{lis1} I. V. Ponomarev, L. I. Deych, and A. A. Lisyansky, Appl.
Phys. Lett. \textbf{85}, 2496 (2004)

\bibitem{izrailev1} F. M. Izrailev and N. M. Makarov, Anomalous Transport in
Low-Dimensional Systems with Correlated Disorder, cond-mat/0507629, July 2005
\end{thebibliography}
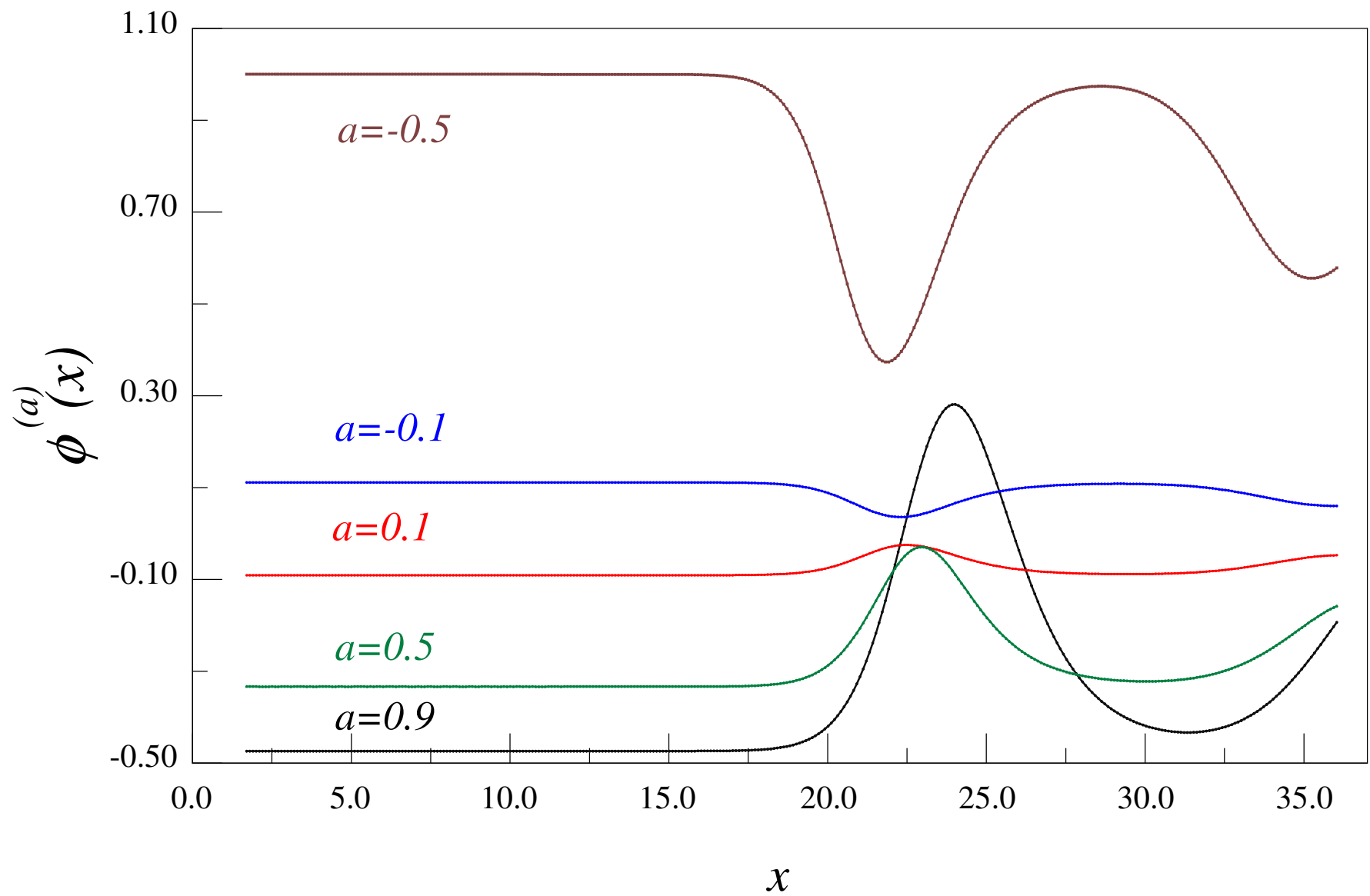
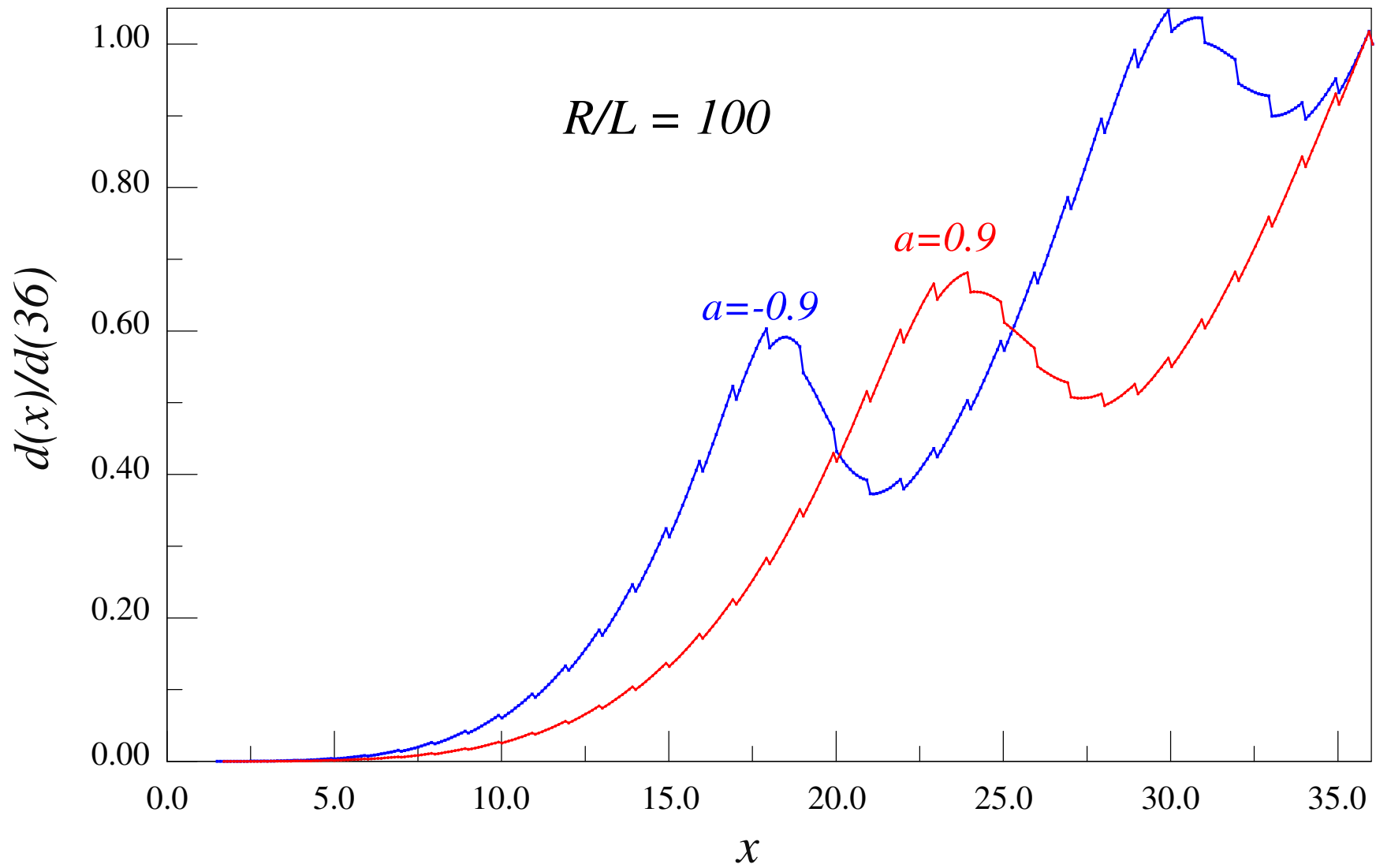
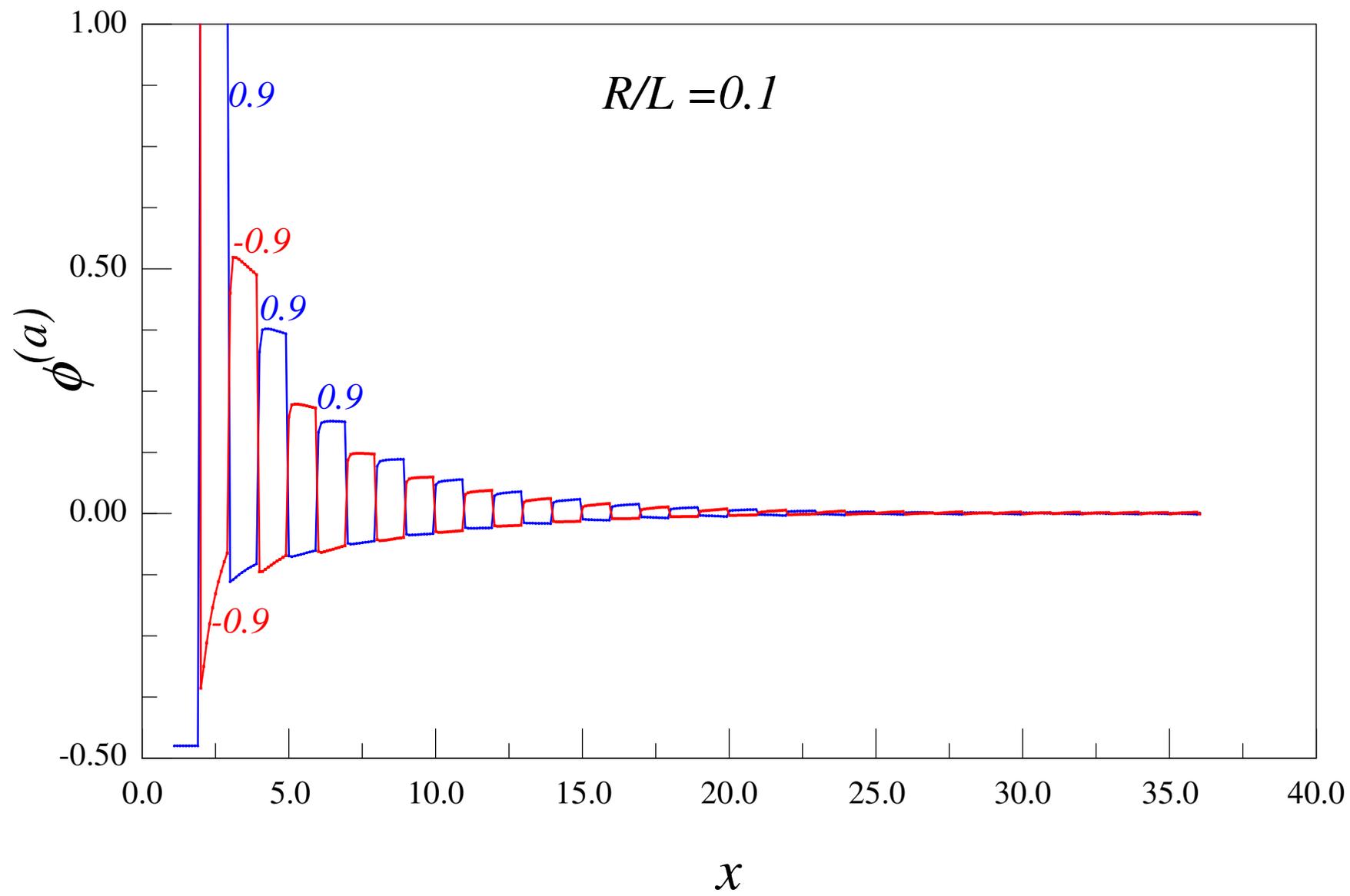
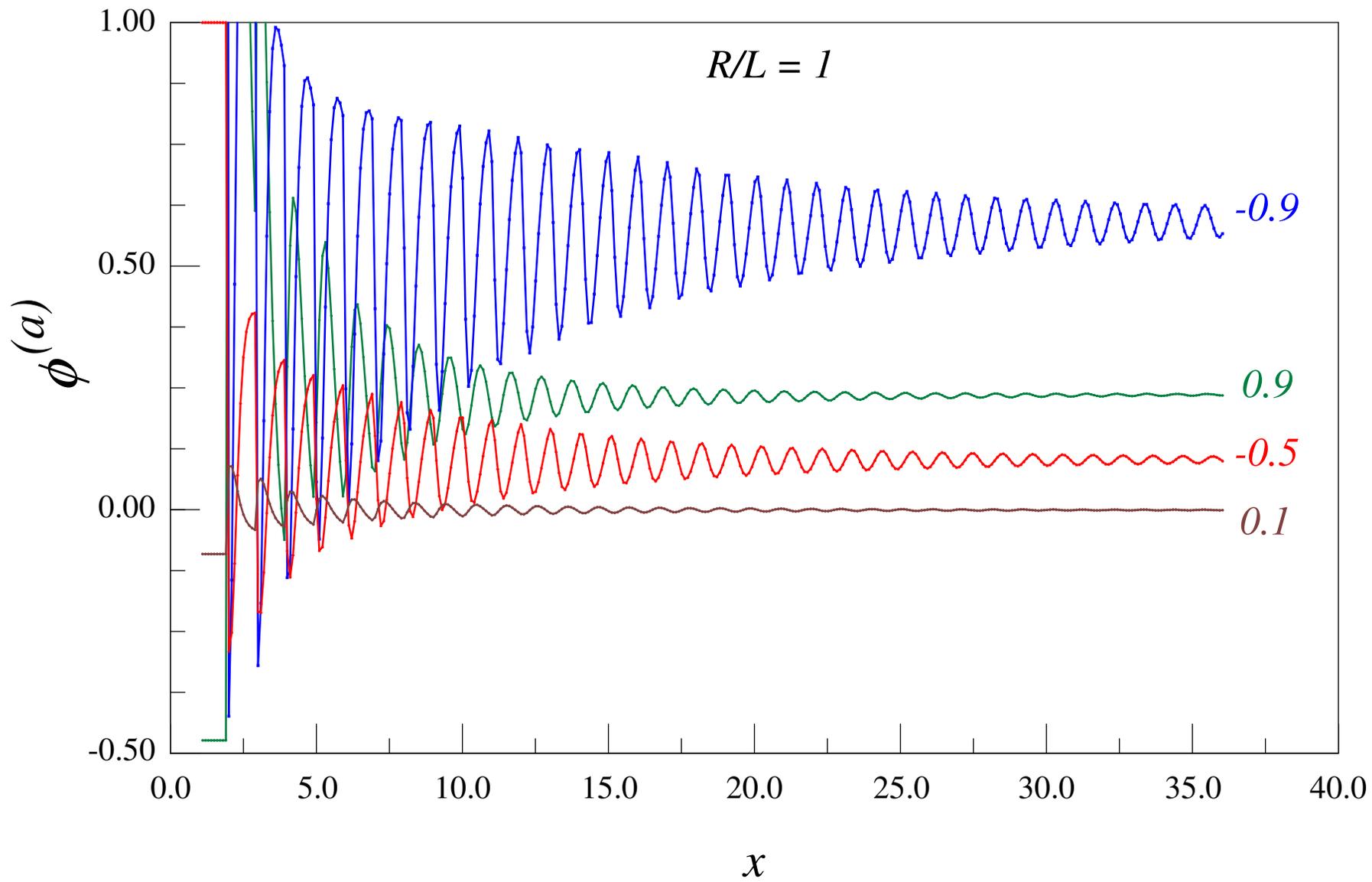
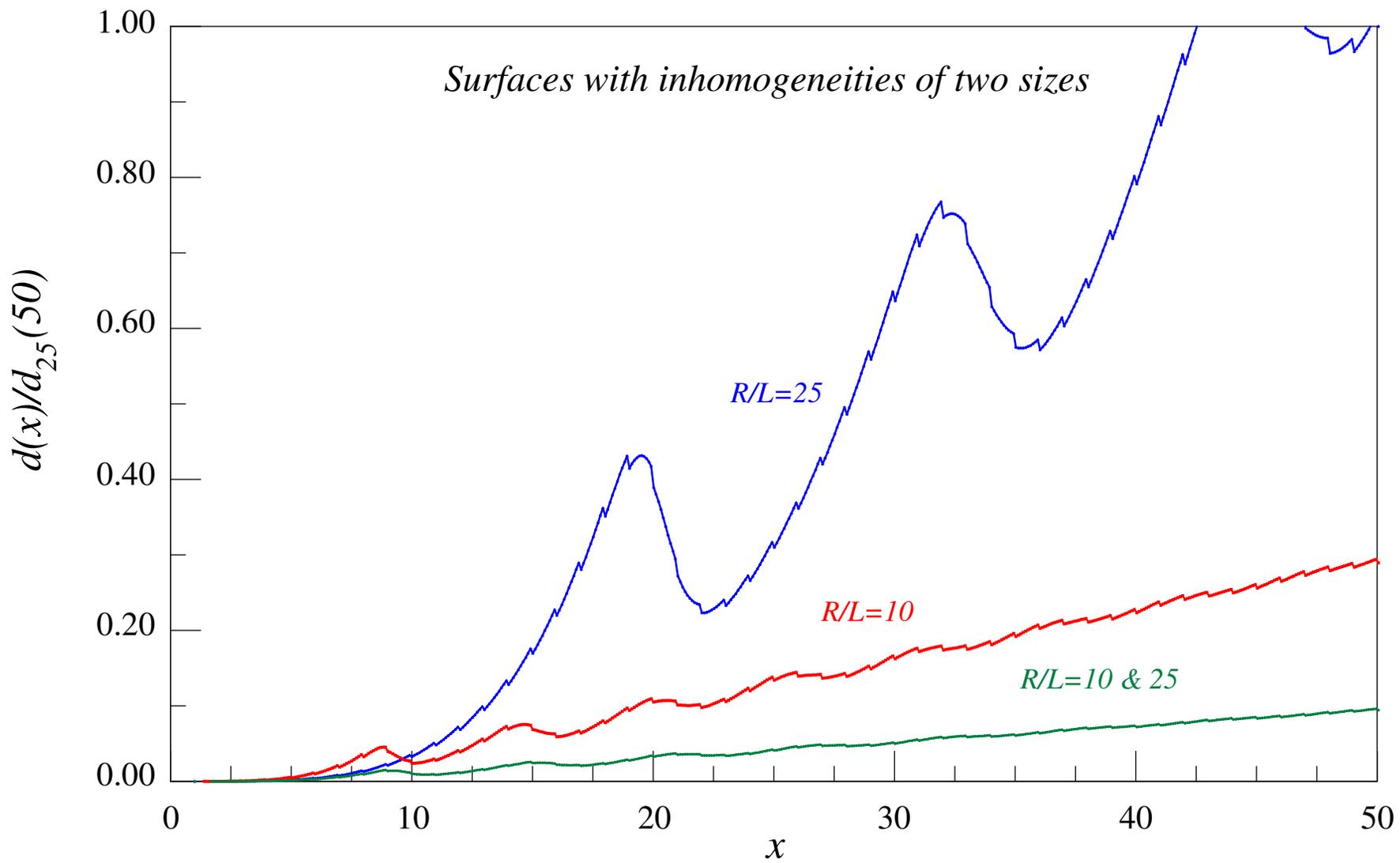
\end{document}